# Enhanced photocatalytic activity of plasmonic Au nanoparticles incorporated $MoS_2$ nanosheets for degradation of organic dyes

Anjali Rani[1], Arun Singh Patel[2], Anirban Chakraborti[3, a)], Kulvinder Singh[4] and Prianka Sharma[1]

[1]Department of Physics, School of Basic & Applied Sciences, Maharaja Agrasen University, H.P., INDIA

[2]Department of Physics, Hindu College, Delhi University, Delhi-110007, INDIA

[3]School of Computational and Integrative Sciences, Jawaharlal Nehru University, New Delhi-110067, INDIA

[4]Department of Chemistry, School of Basic & Applied Sciences, Maharaja Agrasen University, H.P., INDIA

**ABSTRACT:** In the present paper, we investigate the effect of plasmonic Au nanoparticles (NPs) decoration on the photocatalytic efficiency of $MoS_2$ nanosheets. The Au NPs are grown on the surface of chemically exfoliated $MoS_2$ nanosheets by chemical reduction method. Au-$MoS_2$ nanostructures (NSs) are characterized by X-ray diffractometer, Raman spectrometer, absorption spectrophotometer, and transmission electron microscopy. Exfoliated $MoS_2$ and Au-$MoS_2$ NSs are used to study the photocatalytic degradation of organic dyes methyl red (MR) and methylene blue (MB). Under UV-Visible light irradiation, pristine $MoS_2$ shows photo degradation efficiencies between 30% to 46.9% for MR and 23.3% to 44% for MB, with varying exposure times from 30 to 120 min, respectively. However, Au-$MoS_2$ NSs with maximum Au NPs concentration show enhanced degradation efficiency from 70.2 to 96.7% for MR, and from 65.2 to 94.3% for MB. The manifold enhancement of degradation efficiency for both the dyes with Au-$MoS_2$ NSs may be attributed to the presence of Au NPs acting as charge trapping sites in the NSs. We believe this study would have potential application in battling the ill-effects of environmental degradation, which poses a major threat to humans as well as biodiversity.

## I. INTRODUCTION:

With the advent of new technologies, synthetic dyes are excessively used in textile, leather and other industries. These dyes contain organic contaminants that are toxic in nature, and are not completely degraded before releasing them as effluents in wastewater, thereby increasing toxicity of water bodies leading to harmful effects on humans and biodiversity. Photocatalysis is an advanced oxidative process (AOP), which leads to quick formation of hydroxyl radicals that has the capability of oxidizing wide range of pollutants present in water without any preference of selectivity.[1,2] However, if complete mineralization of these contaminants or organic pollutants is not possible, then this AOP at least converts them into

[a)] Corresponding author: Anirban Chakraborti
 E-mail: anirban@jnu.ac.in

intermediate compounds, which are further biodegradable. Thus, to tackle the menace of environmental degradation, it is necessary to develop photocatalysts which are non-toxic, potentially stable and cost-effective for large scale degradation of organic contaminants present in water.[3] Traditionally, semiconductor mediated photocatalysts such as $TiO_2$, ZnO, ZnS, $SnO_2$, CdS, CuS, etc. [4–8] have been studied widely, for complete mineralization of organic pollutants. These semiconductors show high responsivity towards photocatalytic process.[9,10] However, fast recombination of electron-hole pairs in semiconductors, and their limited responsivity towards UV excitation, restrict their application as good photocatalysts.[2,11–13] In a typical photocatalytic process, excitation by photons creates charge separation, by causing the electrons to overcome the band gap of the material. An ideal photocatalyst requires increased surface area for interfacial charge transfer, tailored visible light absorption, tunable band gap, enhanced defect sites for trapping of charge carriers, etc., to enhance the efficiency of the photocatalytic process.[1,11] Thus, the need of the hour is an alternative candidate for a photocatalyst, which can overcome the limitations of pristine semiconductor photocatalysts and can also show improved degradation capabilities.

In recent years, $MoS_2$, a well-known layered transition metal dichalcogenide (TMDC), has gained significant interest in the field of photocatalysis. $MoS_2$ exhibits high chemical reactivity, varied optical properties, enhanced surface area, charge carrier mobility and strong absorption of visible light.[9,14] Its varied band gap from 1.3 eV in bulk to 1.9 eV in monolayer, exhibits an indirect to direct transition, making itself flexibly suitable for light absorption in wide spectrum.[15] Unlike multi-layered materials, presence of tightly bound excitons in monolayer $MoS_2$ produces favourable light absorption properties within direct band gap, making it a visible light assisted photocatalyst.[16–18] However, monolayer $MoS_2$ has very low optical cross-section, lesser number of active sites and faster electron-hole recombination rates, which hinder extensive light-matter interactions leading to weak absorption. In order to overcome faster recombination rate, $MoS_2$ is incorporated with other semiconducting or noble metal nanoparticles. The incorporated materials provide an additional path to the photogenerated electrons and holes, which hinders the recombination rate and enhances the photocatalytic efficiency. Similar to noble metals like Ag, Pd, Pt, etc., Au NPs enjoy the advantageous position due to its non-toxic nature and excellent stability. Au NPs are formed on defect sites of $MoS_2$ and localized by non-covalent bonds. The Au NPs act as a p-type dopant in the $MoS_2$ layer. This leads to strong chemical bonding along the interplanar directions acting as spacer between interlayer $MoS_2$, thereby prohibiting restacking of layers.[11,19–21] Deposition of Au NPs on $MoS_2$ nanosheets serves as sink for electrons, thus hindering the electron-hole recombination and facilitates interfacial transfer of electrons. This increases stability of resulting nano-composites simply by enhancing charge transportation between Au NPs and $MoS_2$.

There are various studies on the growth of metal nanoparticles on $MoS_2$ and their applications in sensing and photovoltaic devices. But there are very few studies on the photocatalytic application of metal, especially Au NPs decorated $MoS_2$ nanostructures.[4,11,22,23] In the present paper, we investigate the effect of plasmonic Au nanoparticles decoration on the photocatalytic efficiency of $MoS_2$ nanosheets. The Au NPs are grown on the surface of chemically exfoliated $MoS_2$ nanosheets by solvent-surfactant assisted liquid exfoliation chemical reduction method. N-Methyl 2-pyrrolidone (NMP) is used as a reducing agent for rapid reduction of metal ion ($Au^{3+}$) to zero-valent gold metal ($Au^0$) in a very short time period. Sodium dodecyl benzene sulfonate (SDBS) is used as surfactant to provide stable dispersions.[24] This solvent-surfactant assisted sonication method helps in avoiding restacking of exfoliated sheets that remain stable over many weeks. These nanostructures are characterized by XRD, Raman spectrometer, UV-Vis absorption spectrophotometer and transmission electron microscopy and are used towards studying the photocatalytic degradation of organic dyes− methyl red (MR) and methylene blue (MB).

## II. EXPERIMENTAL DETAILS:

Molybdenum disulphide ($MoS_2$) powder, n-methyl 2- pyrrolidone (NMP), isopropyl alcohol (IPA), hydrogen tetrachloroaurate trihydrate ($HAuCl_4.3H_2O$), sodium dodecyl benzene sulfonate (SDBS), methyl red and methylene blue dyes were obtained from Sigma Aldrich. All the chemicals were used without further purification. Double distilled water was used as solvent at various stages.

### A. Chemical exfoliation of $MoS_2$ nanosheets:

Synthesis of $MoS_2$ nanosheets have been done via liquid-phase exfoliation method. For the synthesis of $MoS_2$ nanosheets, 20 mL of NMP, IPA and distilled water (volume ratio 3:1:1), a mixed solvent was taken and 0.05 g of bulk $MoS_2$ powder was dispersed in it. This mixture was ultrasonicated at room temperature continuously for 24 hours. After 24 hours of sonication, solution colour changes from dark greyish black to greenish. The resultant green dispersion obtained was retained for 30 min for settling down the un-exfoliated MoS$_2$. Further, 18 mL of clear supernatant was collected and five different sets each of 2 mL of supernatant were taken out and 10 mg of SDBS was added in each set. The solution was then allowed to settle down for the next 30 min.

**B. Synthesis of Au-MoS$_2$ nanostructures:**

For the synthesis of Au-MoS$_2$ nanostructures, a stock solution of 10 mM Au salt (HAuCl$_4$.3H$_2$O) was prepared. Four sets each of 2 mL dispersion of exfoliated MoS$_2$ nanosheets were taken for synthesis of Au-MoS$_2$ nanostructures. In these four sets different quantities of Au salt solution (2$\mu$L, 4$\mu$L, 6$\mu$L and 8$\mu$L) were added and the mixtures were kept for aging for 48 hours, these samples were named SET-I, SET-II, SET-III and SET-IV, respectively. Thus the Au-MoS$_2$ nanostructures with varying concentrations of Au NPs were synthesized.

**C. Characterization techniques:**

The crystalline structure of as-prepared samples were determined by a powder X-ray diffractometer (XRD) with CuKα radiation (λ=0.154 nm) in the 2θ ranges from 20° to 70°. The absorption spectra of MoS$_2$ and Au-MoS$_2$ nanostructures were recorded using absorption spectrophotometer. The absorption spectrophotometer ((Labtronics, LT-2700, India) was used for the photocatalytic study as well. The vibrational modes of MoS$_2$ and the effect of Au NPs on these modes were investigated using a confocal Raman spectrometer having an excitation source as 532 nm laser. The morphologies of MoS$_2$ and Au-MoS$_2$ were analysed by transmission electron microscope JEOL 2100F operated at 200 kV.

**D. Photocatalytic test for dye degradation:**

Methyl red (MR) and methylene blue (MB) are chemically stable and poorly biodegradable azo dye contaminants in wastewater. In this work, we have used these azo dyes (MR and MB) as model reactions to evaluate the photocatalytic activity of as-prepared photocatalysts - MoS$_2$ and Au-MoS$_2$ nanostructures. The dye concentration used in this process was 50 mL of 100 ppm aqueous solution of MR and MB. The dye-solutions were then treated with 1mL of MoS$_2$ and different sets of Au-MoS$_2$ NSs in dark as well as under white light of Xenon lamp (1200 W) for dye degradation for different exposure times (30, 60, 90, and 120 min). After treatment with photocatalysts, the solutions were centrifuged for 10 min at 4000 rpm to remove the residual of the photocatalysts. Thus obtained dye solutions were used to evaluate the photocatalytic properties of the nanostructures by using absorption spectroscopy. In order to evaluate the effect of the exposure time on the photocatalytic efficiency different sets of samples were taken with varying exposure time of 30 min duration.

## III. RESULTS AND DISCUSSION:
### A. X-ray Diffraction:

X-ray diffraction is used to characterize the crystalline structure and to count the microstructures of material in large quantities. Fig. 1 shows the X-ray diffraction patterns of Au, $MoS_2$ and Au-$MoS_2$ NSs.

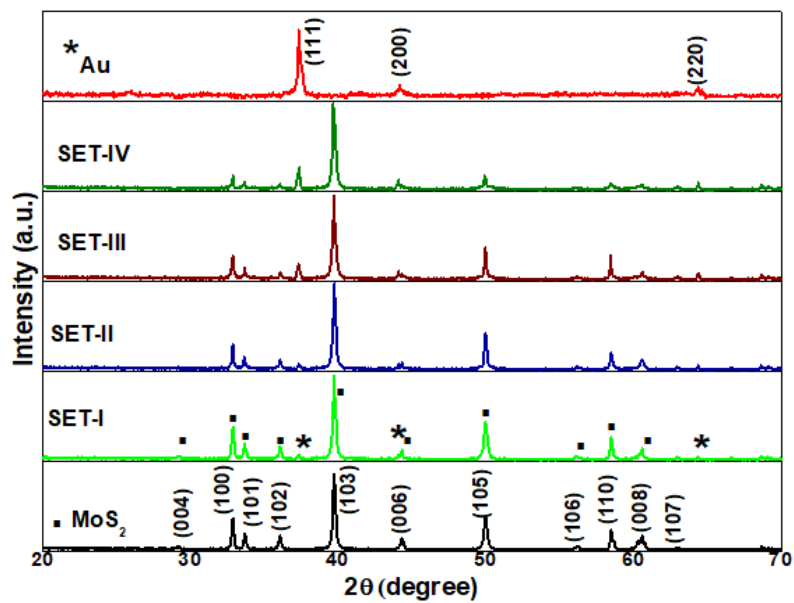

FIG. 1. XRD patterns of the Au, $MoS_2$ and Au-$MoS_2$ nanostructures (SET-I to SET-IV). (asterisk represents Au and black filled square indicates $MoS_2$ XRD patterns).

In these XRD patterns, diffraction peaks at 29.19°, 32.88°, 33.68°, 36.09°, 39.75°, 44.34°, 49.99°, 56.19°, 58.50°, 60.61° and 62.95° are observed which are attributed to the lattice planes (004), (100), (101), (102), (103), (006), (105), (106), (110), (008) and (107) of $MoS_2$ respectively (JCPDS No. 37-1492).[25] After Au NPs decoration, the Au-$MoS_2$ NSs show extra diffraction peaks at 37.34°, 44.17° and 64.40° corresponding to the (111) (200) and (220) planes, respectively of Au phase (JCPDS No. 04-0784)[26] indicating that the FCC Au NPs have been successfully decorated on $MoS_2$ surface. Intensity of Au peak enhances as we increase the concentration of Au, while the intensity of peaks of $MoS_2$ decreases accordingly in SET-I, SET-II, SET-III and SET-IV (with increasing concentration of Au - 2µL, 4µL, 6µL and 8 µL, respectively) which is clearly observed in Fig.1. Lowering in diffraction peak intensity is due to decrease in the exposure area of $MoS_2$ nanosheets in presence of Au NPs. The presence of Au NPs hinders the exposure area of the $MoS_2$. The crystallite size of $MoS_2$ is found to be 44.18 nm in case of pristine while in the SET-I, SET-II, SET-III, and SET-IV it comes out to be 39.01 nm, 37.34 nm, 35.85 nm and 34.88 nm, respectively, which was calculated by using Debye Scherrer's formula (1).

$$D = \frac{k\lambda}{\beta \cos\theta} \tag{1}$$

where $D$ is the mean size of the crystallites, $k$ is a dimensionless shape factor (~0.9), $\lambda$ is the X-ray wavelength ($\lambda$=1.5418 Å), $\beta$ is the line broadening at half the maximum intensity (FWHM) $\theta$ is the position of the diffraction peak. As the dimension decreases, peaks are broadened which shows that the NSs lose their crystallinity with the resultant increase in amorphous to crystalline fraction ratio. Thus, the size dependent broadening and intensity of patterns can be attributed to the reduced crystallite size, in-built strain in the $MoS_2$ lattice due to edge effects, and the loss of the crystalline fraction in small sized NSs.

**B. UV-Visible absorption spectroscopy:**

The optical absorption of pristine $MoS_2$ nanosheets and Au-$MoS_2$ nanostructures has been explored using UV-Vis absorption spectrophotometer. The absorption spectra of $MoS_2$ and Au-$MoS_2$ are shown in Fig. 2. The absorption spectrum of $MoS_2$ nanosheet shows two small humps in the visible range, one at 670 nm and other at 620 nm, which are known as *A* and *B* excitonic peaks, respectively. These peaks arise due to spin-orbit interaction causing splitting of valence band energy levels and the excitonic transitions between splitted valence bands and minima of conduction band at the *K*-point of the Brillouin zone.[27] The interlayer coupling also plays an important role in the valence band splitting. When a small aliquot of gold precursor is added into chemically exfoliated $MoS_2$, a new absorption peak corresponding to the Au plasmon band emerges at around 530 nm, suggesting consumption of $Au^{3+}$ ions and formation of gold nanoparticles. The extent of Au-ion reduction by the spontaneous redox reaction can be estimated by monitoring the absorption peak of $Au^{3+}$ and quantifying the loading level of Au on the surface of $MoS_2$. As we increase the Au concentration, the surface plasmon resonance (SPR) peaks of Au experiences both red shift and increase in the intensity of peaks (as shown in Fig. 2). These shifts infer strong plasmon-excitonic coupling between Au and $MoS_2$. The surface plasmon resonance strongly depends on shape, size and separation of the nanoparticles along with the surrounding environment. In Fig. 2, it is observed that when the concentration of Au is low, the SPR peak is not so prominent, which is due to smaller size and low concentration of Au NPs.

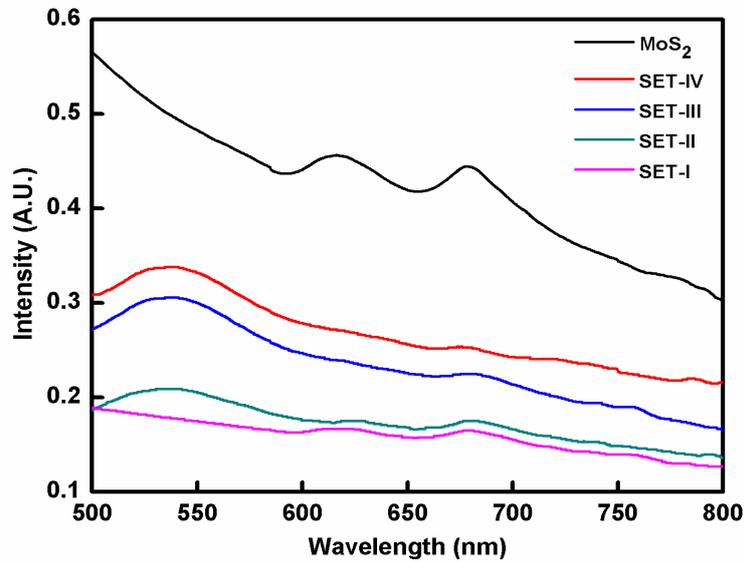

**FIG. 2. Absorption spectra of $MoS_2$ and Au-$MoS_2$ nanostructures with different concentrations of Au.**

## C. Raman Investigations:

Raman spectroscopy has been utilized to investigate the crystallinity and layer thickness of two dimensional (2D) $MoS_2$ in terms of the position and frequency difference of two characteristic vibrational modes, $E^1_{2g}$ and $A_{1g}$. It is also used for finding the effect of lattice strain, doping levels, and the van der Waals interaction at the interface of 2D crystals.[28] The $E^1_{2g}$ mode is attributed to the in-plane vibration of Mo and S atoms and this mode is sensitive to the built-in strain of 2D $MoS_2$, while $A_{1g}$ mode is related to the out-of-plane vibration of S atoms which is a reflection of interlayer van der Waals interaction.

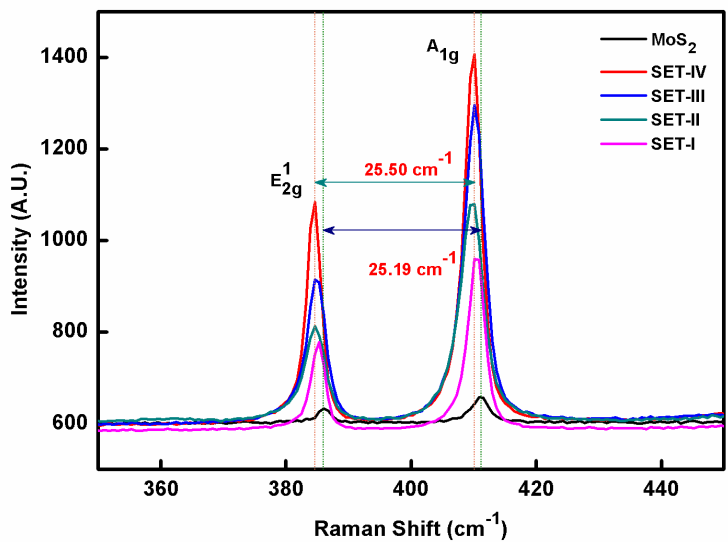

**FIG. 3. Raman spectra of MoS$_2$ and Au-MoS$_2$ nanostructures.**

The Raman spectra of MoS$_2$ shown in Fig. 3 exhibit two vibrational modes centred at 386.03 cm$^{-1}$ and 411.22 cm$^{-1}$. In the case of Au-MoS$_2$ NSs, red shift in of E$^1_{2g}$ and A$_{1g}$ modes is observed. The SET-IV, with maximum loading of Au NPs, exhibits red shift of E$^1_{2g}$ and A$_{1g}$ modes; these peaks are observed at 384.64 cm$^{-1}$ (shift ~ 1.39 cm$^{-1}$) and 410.14 cm$^{-1}$ (shift ~1.08 cm$^{-1}$), respectively. The shifting is attributed to the effect of lattice strain due to curvature of the MoS$_2$ shell. The frequency difference of E$^1_{2g}$ and A$_{1g}$ peaks for MoS$_2$ comes out to be about 25.19 cm$^{-1}$ which renders exfoliation of few-layers MoS$_2$ sheets.[15] Similar frequency difference was observed for Au-MoS$_2$ sheets i.e., 25.50 cm$^{-1}$ for SET-IV. Besides variation in Raman frequency, we have also observed a significant enhancement of Raman peak intensity in the Au-MoS$_2$ nanostructures. This can be attributed to the effect of localized surface plasmon resonance (LSPR) of Au nanoparticle cores, typically called surface enhanced Raman scattering (SERS).[29] The interaction of the incident light with Au NPs excites localized surface plasmons. When there is resonance between the frequency of plasmon oscillation and irradiation, a strong electromagnetic field is formed on the surface. This electromagnetic field leads to a significant increase in the intensity of Raman mode as shown in Fig. 3. The intensity ratio of A$_{1g}$/ E$^1_{2g}$, for exfoliated MoS$_2$ is found to be 1.04. However with the incorporation of Au NPs, the peak intensity ratio increases from 1.23 (SET-I) to 1.30 (SET-IV).

**D. TEM Imaging:**

The shape and size of $MoS_2$ and Au-$MoS_2$ nanostructures were investigated using transmission electron microscopy (TEM). The TEM images of $MoS_2$ and Au-$MoS_2$ nanostructures are shown in Fig. 4.

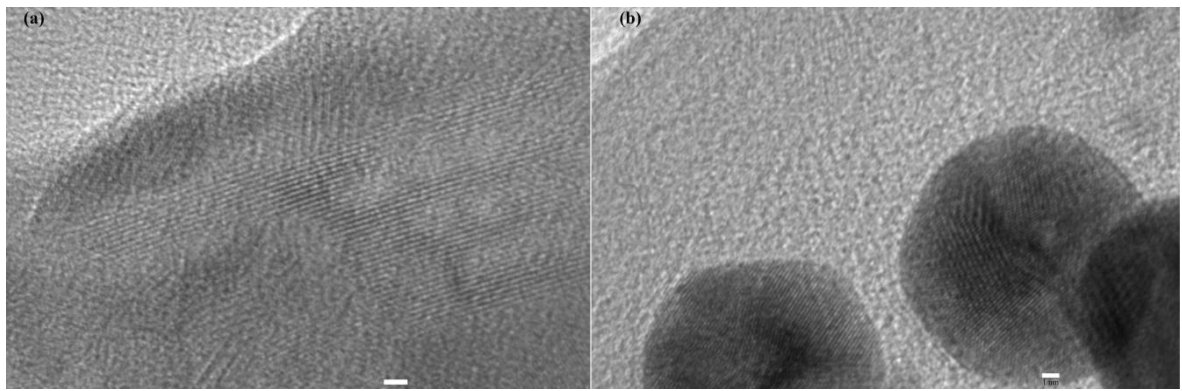

FIG. 4. TEM images of (a) MoS2 and (b) Au-$MoS_2$ nanostructures; the white scale bar is equivalent to 1 nm.

In Fig. 4(a) sheet-like structure is observed, which corresponds to the $MoS_2$. In $MoS_2$ the lattice spacing is of the order of 0.27 nm, which corresponds to (100) plane. In case of Au-$MoS_2$, the Au nanoparticles are found to be of spherical shape and the size of these particles is around 20 nm (Fig. 4(b)). The lattice plane spacing in the Au nanoparticles corresponding to FCC lattice of Au is observed and the spacing is of the order of 0.22 nm, which corresponds to (111) plane of gold lattice.

### E. Photocatalytic Activity of $MoS_2$ and Au-$MoS_2$ nanostructures:

The as-synthesized $MoS_2$ and Au-$MoS_2$ NSs were used for photocatalytic degradation of dye molecules. A blank test in the absence of catalysts was done and negligible self-degradation of dyes was observed.

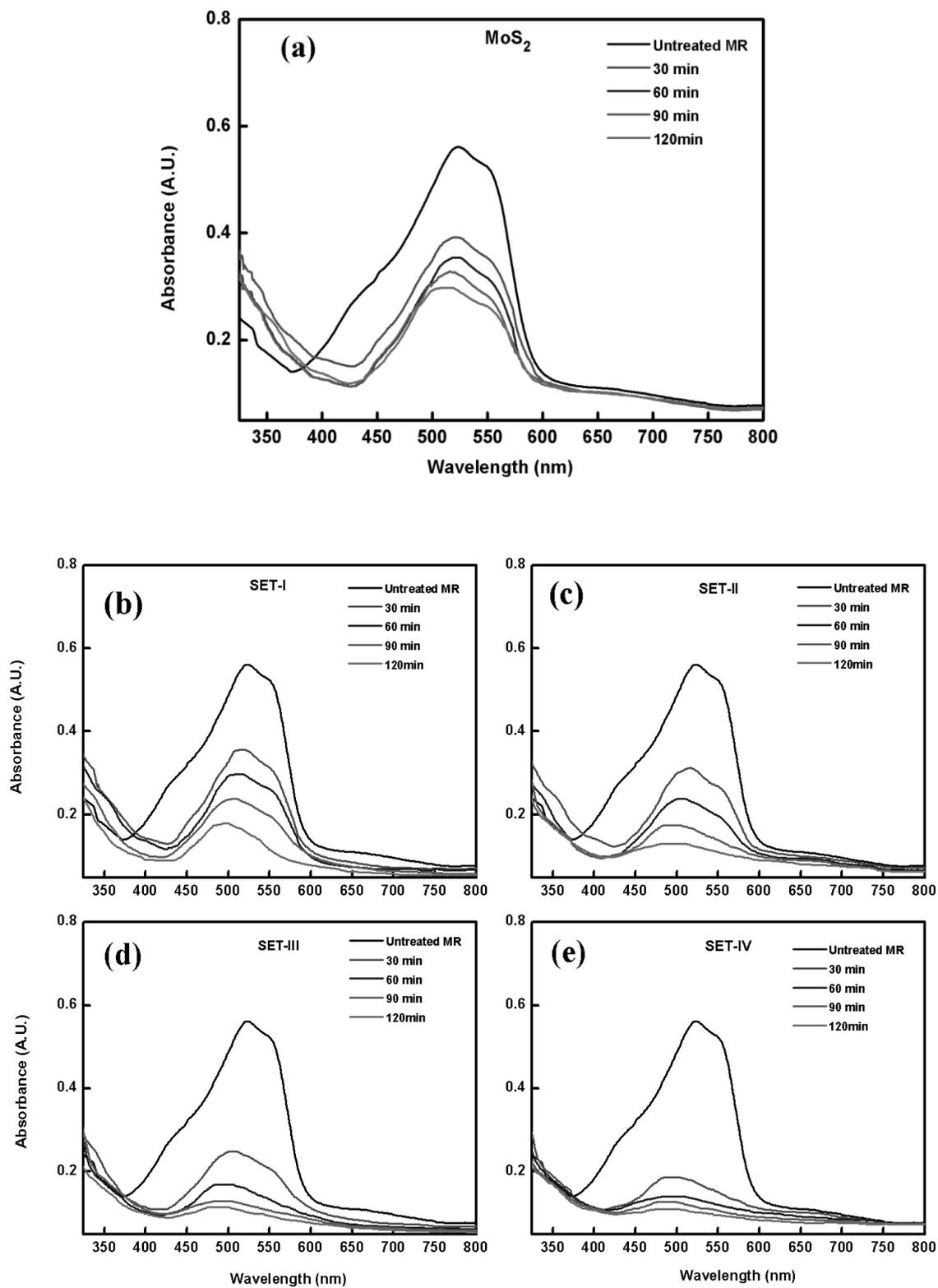

FIG. 5. Absorption spectra of MR with (a) MoS$_2$ nanosheets, (b) SET-I, (c) SET-II, (d) SET-III and (e) SET-IV ( SET I-SET IV, increasing concentration of Au from 2 µL to 8 µL with the difference of 2µL).

The characteristic absorption peak of MR appears at 523 nm. Fig. 5(a)-(e) show the absorption spectra of MR dye molecules under UV-Vis light irradiation in presence of $MoS_2$, SET-I, SET-II, SET-III and SET-IV for different exposure time (30 -120 min). A systematic decrease in the absorbance of the dye molecules is observed with increasing exposure time in presence of $MoS_2$ (Fig. 5(a)). This is due to the degradation of MR dye molecules. The degradation is attributed to the electron-hole pair generation in the $MoS_2$ nanosheets. These electron-hole pairs interact with oxygen and water molecules and create reactive oxygen species which interact with the dye molecules and decompose them. In case of pristine $MoS_2$, photo degradation efficiency is limited due to the direct band gap of $MoS_2$. The direct bandgap causes fast recombination of photogenerated electron-hole pairs in $MoS_2$. The degradation efficiency of $MoS_2$ can be enhanced by incorporating Au nanoparticles in the $MoS_2$. In Fig. 5(b)-(e) the characteristic peak of MR shows downshift (hypochromic shift) as the concentration of Au increases in Au-$MoS_2$ NSs. SET-IV exhibits maximum decrease in the absorbance of the dye molecules. The concentration of dye was monitored with the aid of absorption spectroscopy. According to Beer-Lambert's law, absorbance is proportional to the concentration of dye molecules, so its percentage degradation efficiency can be calculated by following equation (2).[20]

$$R = \frac{C_o - C}{C_o} \times 100\% = \frac{A_o - A}{A_o} \times 100\% \tag{2}$$

where, $C_o$, $C$ are concentration and $A_o$, $A$ are absorbance of dye at initial time and after final time $t$, respectively.

For quantitative understanding of the effect of photocatalysts on reaction kinetics of the dyes, we have utilized the pseudo-first order reaction mechanism model. The rate constant $k$ of the degradation was investigated using equation (3),

$$\ln\ln\left(\frac{C}{C_0}\right) = -kt \tag{3}$$

where, $C_o$ is initial concentration and $C$ is concentration at $t$ exposure time, $k$ is the pseudo-first-order rate constant. The rate constants are obtained from the regression line between $\ln(C/C_o)$ and time $t$. Fig 6(a) represents the percentage degradation of MR with $MoS_2$, SET-I, SET-II, SET-III and SET-IV as photocatalysts. It is found that degradation efficiency of $MoS_2$ enhances from 30 to 46.9%, with varying the exposure time from 30 min to 120 min.

For SET-I the degradation efficiency enhances from 37 to 68%, for SET-II it enhances from 45 to 80.2%, for SET-III from 55 to 89.3% and for SET-IV the degradation efficiency enhances from 70.2 to 96.7% after irradiating with UV-Vis light and varying the exposure time from 30 to 120 minute. Thus, we observe that with increase in Au concentration from SET-I to SET-IV and varying the exposure time from 30 min to 120 min, the degradation efficiency increases. For longer exposure to UV-Vis light irradiation (120 min), more electron-hole pairs are generated, which causes enhanced degradation of dye molecules.

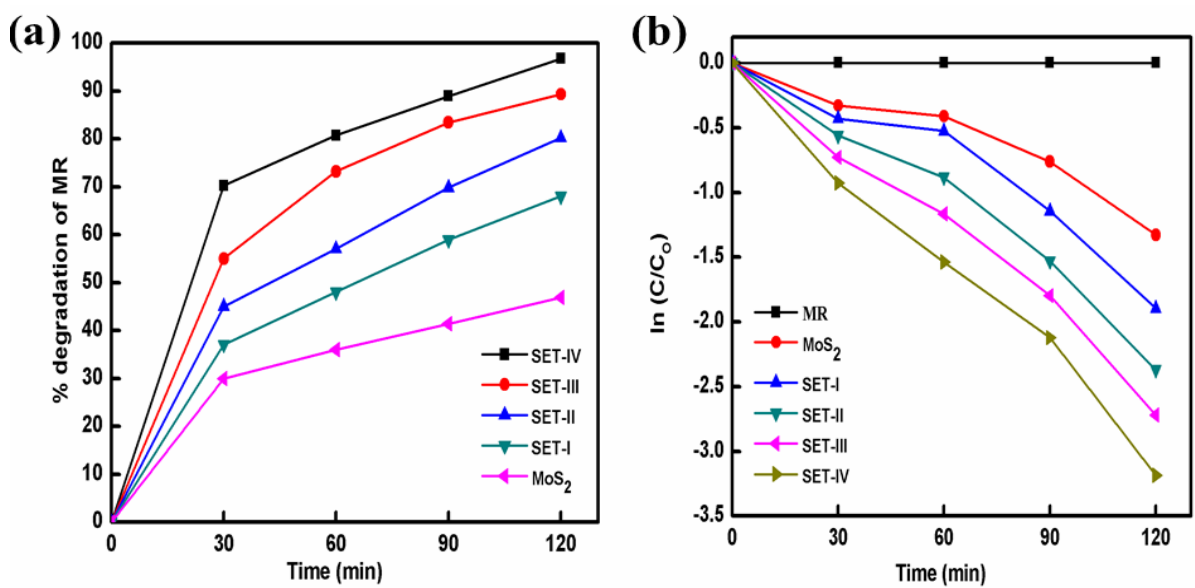

FIG. 6(a) Percentage degradation of MR with $MoS_2$, SET-I, SET-II, SET-III and SET-IV, (b) kinetic graph of MR with photocatalysts $MoS_2$ and different sets of Au-$MoS_2$.

The reaction kinetics corresponding to photocatalytic degradation of MR are plotted in Fig. 6(b). The rate constant values calculated from the regression lines of MR for $MoS_2$, SET-I, SET-II, SET-III and SET-IV are -0.5660 $min^{-1}$, -0.7996 $min^{-1}$, -1.0693 $min^{-1}$, -1.2832 $min^{-1}$ and -1.5551 $min^{-1}$, respectively. This enhancement in mod values of rate constant of the photocatalysts for MR signifies major degradation of dye by SET-IV as compared to other photocatalysts. The % degradation of dye by $MoS_2$, SET-I, SET-II, SET-III and SET-IV along with rate constant values for 120 min of irradiation time are compiled in Table 1.

**TABLE 1 Percentage degradation of MR with MoS$_2$, SET-I, SET-II, SET-III and SET-IV along with rate constants for 120 min of irradiation time.**

| Name of Samples | % Degradation of MR with irradiation time | | | | Rate Constants (min$^{-1}$) |
|---|---|---|---|---|---|
| | 30 min | 60 min | 90 min | 120 min | |
| MoS$_2$ | 30 | 36 | 41.4 | 46.9 | -0.5660 |
| SET-I | 37 | 48 | 58.9 | 68 | -0.7996 |
| SET-II | 45 | 57 | 69.8 | 80.2 | -1.0693 |
| SET-III | 55 | 73.2 | 83.4 | 89.3 | -1.2832 |
| SET-IV | 70.2 | 80.7 | 88.9 | 96.7 | -1.5551 |

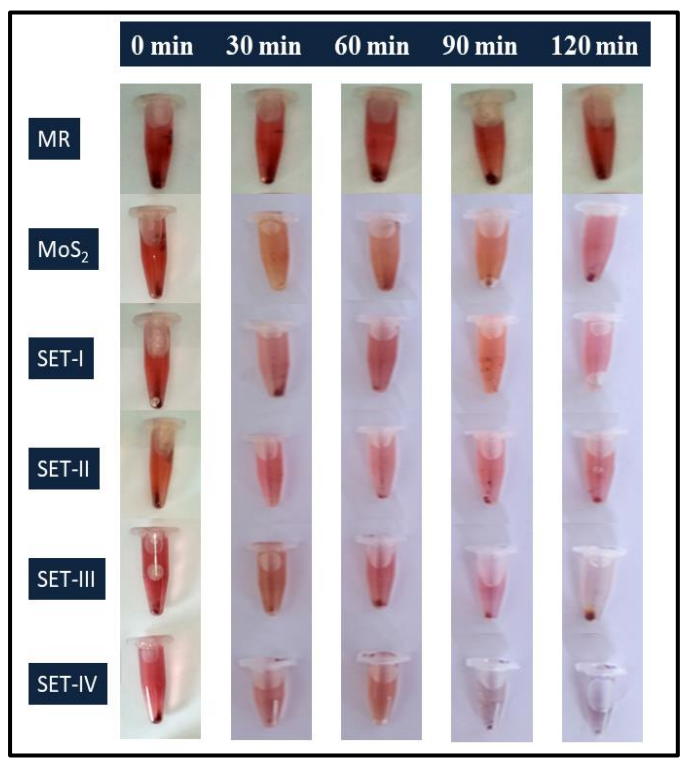

FIG. 7. Pictorial representation of MR degradation in 0, 30, 60, 90, and 120 minutes with MoS$_2$, SET-I, SET-II, SET-III and SET-IV (SET-I to SET-IV increasing conc. of Au in MoS$_2$ by the factor 2μL).

A pictorial image of MR dye under different photocatalytic treatment is presented in Fig. 7. From the image, it is clearly observed that increasing the Au NPs concentration and the exposure time of UV-Vis light irradiation, the colour of dye fades out, which reflects degradation of the dye molecules.

Similar to the MR dye molecules, we have also investigated the photodegradation of MB dye using $MoS_2$ and different sets of Au-$MoS_2$ NSs as catalysts under UV-Visible light irradiation. The characteristic peak of MB is monitored at 663 nm by using a UV-Visible spectrophotometer. In Fig. 8(a), with $MoS_2$ as photocatalyst, a downshift in the absorbance curve is observed with increase in the irradiation time. The change in absorbance is moderate, which indicates less efficiency towards dye decolorization due to confined band gap of $MoS_2$. Whereas when different sets of Au-$MoS_2$ NSs are used as photocatalysts, as shown in Fig. 8(b)-(e), a noticeable decrease in absorbance with time is observed. From the absorption data it is clear that coupling of Au with $MoS_2$, especially SET-IV among all four sets provides better degradation efficiencies as compared to $MoS_2$ and other sets of Au-$MoS_2$.

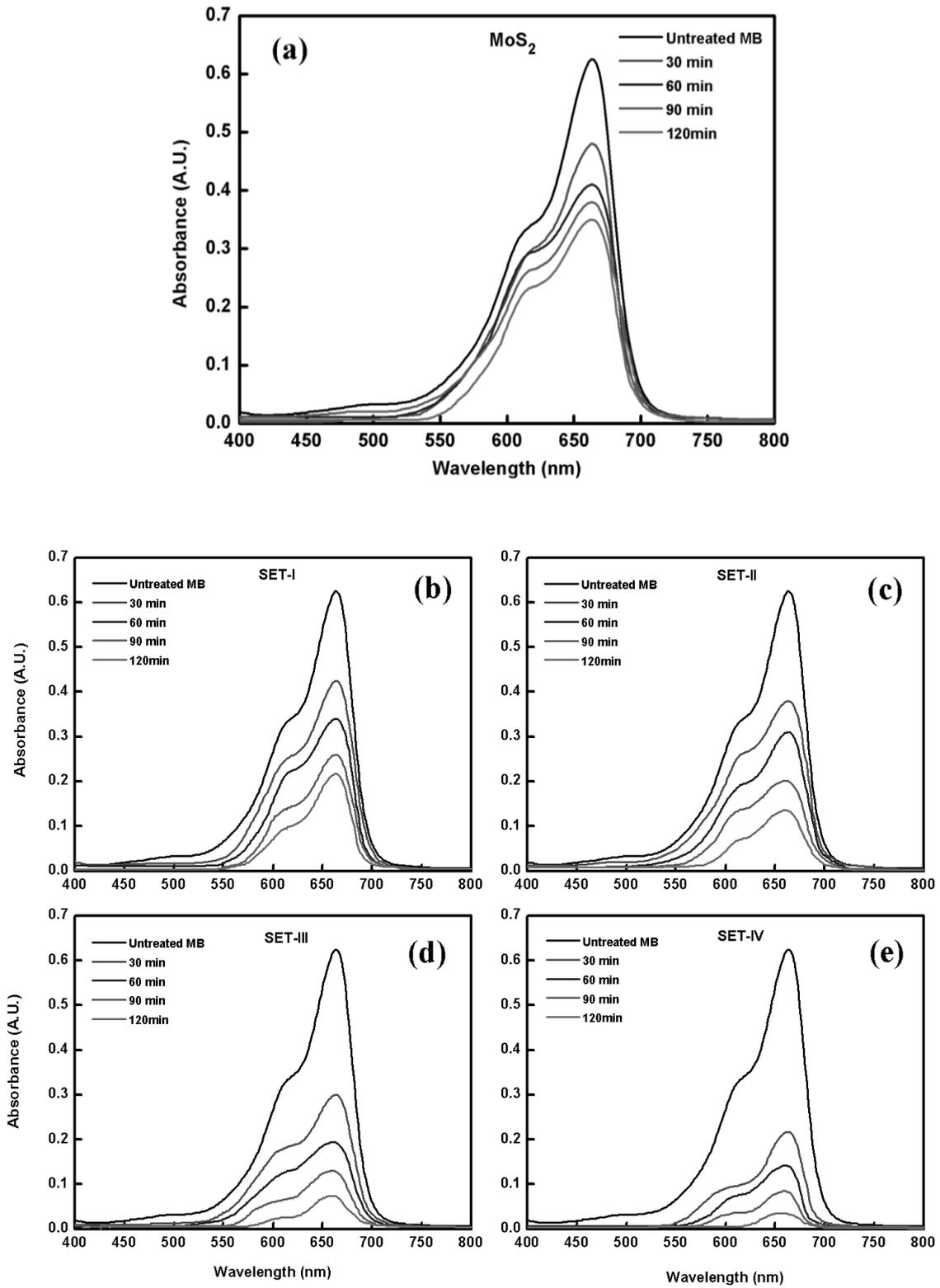

FIG. 8. Absorption spectra of MB after exposure of UV-Vis light for different time in presence of (a) $MoS_2$ nanostructures, (b) SET-I, (c) SET-II, (d) SET-III and (e) SET-IV.

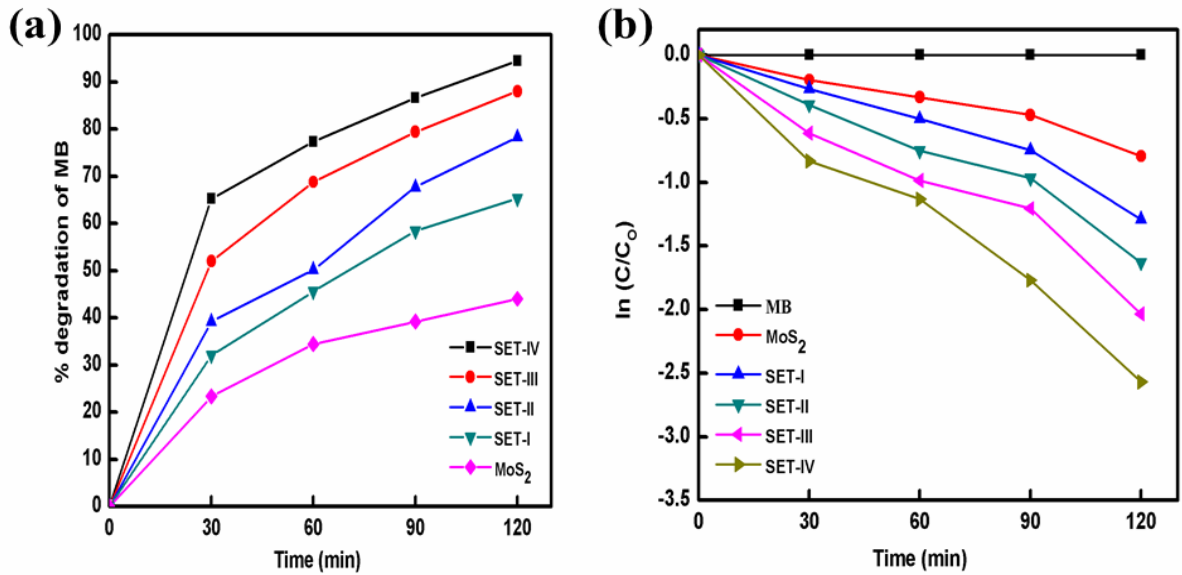

FIG. 9(a) Percentage degradation of MB with $MoS_2$, SET-I, SET-II, SET-III and SET-IV, (b) Kinetic graph of MB with photocatalysts $MoS_2$ and different sets of Au-$MoS_2$.

To examine the percentage degradation of dye by the catalysts, equation (2) is applied and the results are shown in Fig. 9(a). Fig. 9(a) illustrates that percentage degradation efficiency increases with different catalysts, for $MoS_2$ it increases from 23.3 to 44%; for SET-I from 32 to 65.2%; for SET-II from 39.2 to 78.2%; for SET-III from 52 to 88%; and for SET-IV the efficiency increases from 65.2 to 94.3% with exposure time varying from 30 to 120 min. These results of percentage degradation confirm that the SET-IV exhibits much better degradation efficiency as compared to other catalysts. The reaction kinetics corresponding to photocatalytic degradation of MB are plotted in Fig. 9(b). The rate constant values obtained from the regression lines of MB are -0.3587 $min^{-1}$ for $MoS_2$, -0.5612 $min^{-1}$ for SET-I, -0.7490 $min^{-1}$ for SET-II, -0.9689 $min^{-1}$ for SET-III, and -1.2614 $min^{-1}$ for SET-IV. The increase in mod value of rate constants of photocatalysts for MB indicates major elimination of dye by SET-IV. The percentage degradation of dye with different photocatalysts ($MoS_2$, SET-I, SET-II, SET-III and SET-IV) and consequent rate constant values are compiled in Table 2.

**TABLE 2 Percentage degradation of MB with MoS$_2$, SET-I, SET-II, SET-III and SET-IV along with rate constants for 120 min of irradiation time.**

| Name of Samples | % Degradation of MB with irradiation time | | | | Rate Constants (min$^{-1}$) |
|---|---|---|---|---|---|
| | 30 min | 60 min | 90 min | 120 min | |
| MoS$_2$ | 23.3 | 34.4 | 39.2 | 44 | -0.3587 |
| SET-I | 32 | 45.6 | 58.4 | 65.2 | -0.5612 |
| SET-II | 39.2 | 50.4 | 67.4 | 78.2 | -0.7490 |
| SET-III | 52 | 68.8 | 79.2 | 88 | -0.9689 |
| SET-IV | 65.2 | 77.2 | 86.4 | 94.3 | -1.2614 |

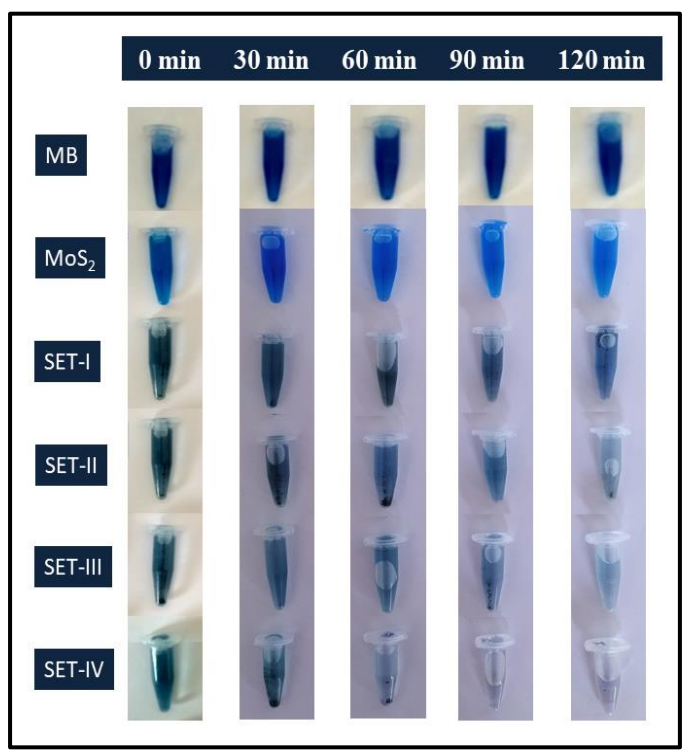

FIG. 10. Pictorial representation of MB degradation in 120 minutes with different photocatalysts (MoS$_2$, SET-I, SET-II, SET-III and SET-IV; SET-I to SET-IV increasing concentration of Au in MoS$_2$ by factor 2μL).

Fig. 10 illustrates the pictorial images of MB degradation with different irradiation time. It is observed that under UV-Visible light irradiation, the dye colour changes. SET IV exhibits maximum fading of colour which correlates with the lower absorption spectra in Fig. 8(e) corresponding to maximum degradation of MB in 120 min.

## IV. PHOTOCATALYTIC MECHANISM:

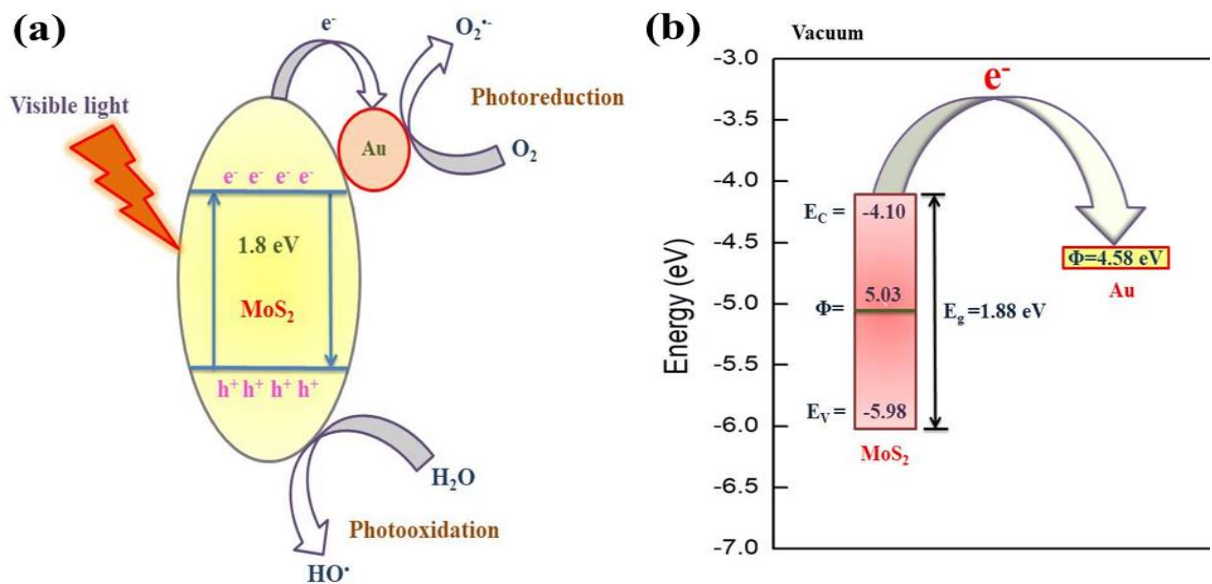

FIG. 11. (a) Charge transfer mechanism in Au-MoS$_2$ nanostructures and (b) energy level diagram of the Au-MoS$_2$ interface.

Due to enlarged surface area, two dimensional MoS$_2$ nanosheets enhance the interfacial charge transfer processes between photocatalyst and oxygen molecules dissolved in wastewater. Basal planes of MoS$_2$ are quite stable under photocatalytic activity but the edge sites or defect sites participate actively in the photocatalysis process. Au NPs nucleate at these highly energetic defect sites such as edges or line defects and enhance the charge transportation between Au NPs and MoS$_2$. Fig. 11(a) shows the mechanism of charge transfer from MoS$_2$ to Au NPs and Fig. 11(b) represents the energy level diagram of Au-MoS$_2$ interface. From Fig. 11(b), the work function of MoS$_2$ is 5.03 eV whereas for gold it is 4.58 eV. The Fermi level of MoS$_2$ is well situated above the reduction potential (1.50 eV) of Au. The electron-hole pairs are created under UV-Visible light irradiation with photon energy greater than the energy band gap ($E_g$) of MoS$_2$ (Eq. 4). However, these electron-hole pairs have a tendency of recombining easily. Incorporation of Au NPs in MoS$_2$ nanosheets helps to provide an additional path to the photogenerated electrons in the conduction band of MoS$_2$ to move on to the surface of Au NPs. This accelerates efficient splitting of charge carriers which helps in minimizing possible electron-hole recombination. These Au NPs act as charge trapping centres. The photoexcited electrons accumulating on the surface of Au NPs and create superoxide radical anions ($O_2^-$) which reacts with oxygen molecules dissolved in wastewater (Eq. 6).[30] With increase in the concentration of Au NPs the charge trapping centres are also increased which help to create more holes ($h^+$) in $MoS_2$ and electrons ($e^-$) on the surface of Au. These electrons and holes act with water and oxygen molecules and create more reactive oxygen species (ROSs). The ROSs helps in photo-reduction of organic dyes (Eq. 7) causing degradation of dye molecules. Thus, increasing the concentration of Au NPs helps to enhance the degradation efficiency. On the other hand, holes in the valence band (VB) of $MoS_2$ react with water molecules to form hydroxyl radicals ($\cdot OH$) which stimulate the degradation of dye molecules (Eq. 8 and Eq. 9) by photo-oxidation process.[31]

$$Au\text{-}MoS_2 + h\nu \rightarrow MoS_2\ (h^+)/\ Au\ (e^-) \tag{4}$$

$$MoS_2\ (h^+) + H_2O \rightarrow \ ^\cdot OH + H^+ \tag{5}$$

$$Au\ (e^-) + O_2 \rightarrow O_2^- \tag{6}$$

$$O_2^- + H_2O \rightarrow \ ^\cdot HO_2 + OH^- \tag{7}$$

$$^\cdot HO_2 + H_2O \rightarrow H_2O_2 + \ ^\cdot OH \tag{8}$$

$$H_2O_2 \rightarrow 2\ ^\cdot OH \tag{9}$$

The noble metal is recognized to act as sink for photo-induced charge carriers promoting interfacial charge transfer processes and hence enhances the photo degradation efficiency of Au-$MoS_2$ nanostructures.[32]

### A. Comparison of degradation of MR and MB:

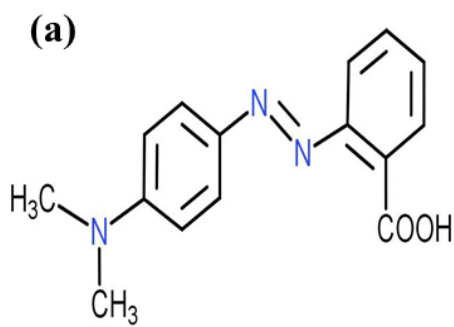
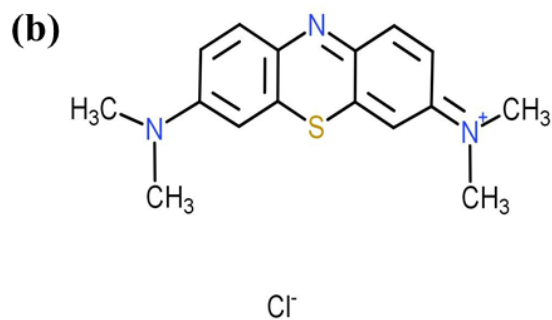

FIG. 12. Molecular structure of (a) methyl red and (b) methylene blue.

Fig. 12(a) and (b) represents the molecular structure of MR and MB, respectively. Colour of any dye depends broadly on the chromophoric sites present in that dye. MR and MB differ not only in their molecular structure but also differ in their functional groups attached and their extent of ionization in any aqueous solution. Both the dyes are basically cationic by nature. Degradation of MR and MB with Au-MoS$_2$ NSs depends upon as many chromophoric sites present in them and their destabilization with free radicals generated by photocatalysis mechanism reaction.[20] MR has 1 chromophoric site while 4 active sites are present in MB. The $^.$OH or $O_2^-$ radicals generated under irradiation by UV-Visible light destabilize or break the chromophoric sites present in MR and MB.[33] Comparing the molecular structure of MR and MB, MB has 4 methyl groups in addition to these chromophoric sites which induces more inductive effect and prevent the dye from degradation easily. This leads to slower degradation efficiency of MB as compared to MR.

## V. CONCLUSION:

In summary, heterostructures of Au-MoS$_2$ nanostructures have been synthesized by the reduction of [AuCl$_4$]$^-$ on the surface of MoS$_2$. The Au decorated MoS$_2$ NSs have been found to improve the efficiency of interfacial charge transfer process in photocatalysis. The photocatalytic studies indicate the decolourization of MR and MB dyes in presence of exfoliated MoS$_2$, and different sets of Au-MoS$_2$ catalysts. It is found that Au decorated MoS$_2$ NSs improve the efficiency towards photodegradation of organic azo dyes MR and MB. The deposited noble metal NPs exhibit interfacial electron transfer which leads to suppression of recombination of charge carriers. This study provides a detailed insight into the fabrication of UV-Visible light active Au-MoS$_2$ nanostructures with enhanced photocatalytic performance and photo-stability. As a result, Au-MoS$_2$ nanostructures act as favourable candidates for photocatalytic degradation of toxic organic contaminants in wastewater using UV-Visible light irradiation, which is beneficial for environmental remediation concern.

**Acknowledgment:** Authors are thankful to AIRF JNU for TEM and Raman measurements.

**References:**

[1] K.C. Lalithambika, K. Shanmugapriya, and S. Sriram, Appl. Phys. A Mater. Sci. Process. **125**, 1 (2019).

[2] Z. Li, X. Meng, and Z. Zhang, J. Photochem. Photobiol. C Photochem. Rev. **35**, 39 (2018).

[3] M. Sharma, P.K. Mohapatra, and D. Bahadur, Front. Mater. Sci. **11**, 366 (2017).

[4] A. Rani, K. Singh, A.S. Patel, A. Chakraborti, S. Kumar, K. Ghosh, and P. Sharma, Chem. Phys. Lett. (2019).

[5] S. Kapatel and C.K. Sumesh, Electron. Mater. Lett. **15**, 119 (2019).

[6] M. Mittal, A. Gupta, and O.P. Pandey, Sol. Energy **165**, 206 (2018).

[7] E. Benavente, F. Durán, C. Sotomayor-Torres, and G. González, J. Phys. Chem. Solids **113**, 119 (2018).

[8] K.M. Reza, A. Kurny, and F. Gulshan, Appl. Water Sci. **7**, 1569 (2017).

[9] K. Zhu, L. Luo, and T. Peng, J. Wuhan Univ. Technol. Mater. Sci. Ed. **34**, 883 (2019).

[10] A. Pal, T.K. Jana, and K. Chatterjee, in *AIP Conf. Proc.* (American Institute of Physics Inc., 2018), p. 050138.

[11] H. Wang, L. Cui, S. Chen, M. Guo, S. Lu, and Y. Xiang, J. Appl. Phys. **126**, 015101 (2019).

[12] A. Bumajdad, M. Madkour, Y. Abdel-Moneam, and M. El-Kemary, J. Mater. Sci. **49**, 1743 (2014).

[13] D. Pan, J. Jiao, Z. Li, Y. Guo, C. Feng, Y. Liu, L. Wang, and M. Wu, ACS Sustain. Chem. Eng. **3**, 2405 (2015).

[14] S. Mouri, Y. Miyauchi, and K. Matsuda, Nano Lett. **13**, 5944 (2013).

[15] T.P. Nguyen, W. Sohn, J.H. Oh, H.W. Jang, and S.Y. Kim, J. Phys. Chem. C **120**, 10078 (2016).

[16] S.V.P. Vattikuti, C. Byon, C.V. Reddy, and R.V.S.S.N. Ravikumar, RSC Adv. **5**, 86675 (2015).

[17] J. Cheng, L. Han, Y. Wei, and Q. Chen, MATEC Web Conf. **108**, 01008 (2017).

[18] S.V.P. Vattikuti, P.C. Nagajyothi, and J. Shim, Mater. Res. Express **5**, 095016 (2018).

[19] M. Sharma, P.K. Mohapatra, and D. Bahadur, Front. Mater. Sci. **11**, 366 (2017).

[20] R. Singh, P.B. Barman, and D. Sharma, J. Mater. Sci. Mater. Electron. **28**, 5705 (2017).

[21] S. Guo, X. Li, J. Zhu, T. Tong, and B. Wei, Small **12**, 5692 (2016).

[22] M. Sigiro, in *AIP Conf. Proc.* (American Institute of Physics Inc., 2017).

[23] M. Velický, G.E. Donnelly, W.R. Hendren, S. McFarland, D. Scullion, W.J.I. Debenedetti, G.C. Correa, Y. Han, A.J. Wain, M.A. Hines, D.A. Muller, K.S. Novoselov, H.D. Abruna, R.M. Bowman, E.J.G. Santos, and F. Huang, ACS Nano **12**, 10463 (2018).

[24] S.H. Lee, D.H. Lee, W.J. Lee, and S.O. Kim, Adv. Funct. Mater. **21**, 1338 (2011).

[25] Z. Wu, D. Wang, and A. Sun, J. Mater. Sci. **45**, 182 (2010).

[26] F. Ochanda and W.E. Jones, Langmuir **21**, 10791 (2005).

[27] M.D.J. Quinn, N.H. Ho, and S.M. Notley, ACS Appl. Mater. Interfaces **5**, 12751 (2013).

[28] Y. Li, J.D. Cain, E.D. Hanson, A.A. Murthy, S. Hao, F. Shi, Q. Li, C. Wolverton, X. Chen, and V.P. Dravid, Nano Lett. **16**, 7696 (2016).

[29] Z. Li, S. Jiang, Y. Huo, M. Liu, C. Yang, C. Zhang, X. Liu, Y. Sheng, C. Li, and B. Man, Opt. Express **24**, 26097 (2016).

[30] V. Kandavelu, H. Kastien, and K. Ravindranathan Thampi, Appl. Catal. B Environ. **48**, 101 (2004).

[31] G. Marcì, V. Augugliaro, M.J. López-Muñoz, C. Martín, L. Palmisano, V. Rives, M. Schiavello, R.J.D. Tilley, and A.M. Venezia, J. Phys. Chem. B **105**, 1033 (2001).

[32] J. Shakya, A.S. Patel, F. Singh, and T. Mohanty, Appl. Phys. Lett. **108**, (2016).

[33] A.R. Khataee and M.B. Kasiri, J. Mol. Catal. A Chem. **328**, 8 (2010).